\def\hst{{\sl HST}}
\def\spitzer{{\sl Spitzer}}

\def\vlt{{\sl VLT}}
\def\irtf{{\sl IRTF}}
\def\gemini{{\sl Gemini}}
\documentclass[12pt,preprint]{aastex}
\usepackage{subfigure}
\usepackage{epsf}
\usepackage{color}
\usepackage{graphicx}
\usepackage{graphics}
\usepackage{epsfig}
\usepackage{longtable}
\usepackage{multirow}
\usepackage[caption=false]{subfig}
\begin{document}
\title{\irtf/TEXES Observations of the \ion{H}{2} Regions H1 and H2  in the Galactic Centre}
\author{Hui Dong$^{1,2}$, John H. Lacy$^3$, Rainer Sch{\"o}del$^1$, Francisco Nogueras-Lara$^1$,Teresa Gallego-Calvente$^1$, Jon Mauerhan$^{4}$, 
  Q. Daniel Wang$^5$, Angela Cotera$^6$,  Eulalia Gallego-Cano$^1$}

\affil{$^1$ Instituto de Astrof\'{i}sica de Andaluc\'{i}a (CSIC), Glorieta de la Astronom\'{i}a S/N, E-18008 Granada, Spain}\affil{$^2$ National Optical Astronomy Observatory,
Tucson, AZ, 85719, USA}\affil{$^3$ Department of Astronomy, University of Texas, Austin, TX 78712, USA}
\affil{$^4$ Department of Astronomy, University of California, Berkeley,
  CA, 94720, USA}
\affil{$^5$ Department of Astronomy, University of Massachusetts,
Amherst, MA, 01003, USA}
\affil{$^6$ SETI Institute, Mountain View, CA, 94043, USA}

\affil{E-mail: hdong@iaa.es}

\begin{abstract}
We present new [\ion{Ne}{2}] (12.8 $\mu$m) \irtf/TEXES observations of the Galactic Center \ion{H}{2} regions
H1 and H2, which are at a projected distance of $\sim$11 pc  
from the center of the Galaxy.  The new observations allow to map the radial velocity distributions of ionized gas. 
The high spectroscopic 
resolution ($\sim$4 {\rm km s$^{-1}$}) helps us to disentangle different velocity components
and enables us to resolve previous ambiguity regarding the nature of these sources. 
The spatial distributions of the intensity and 
radial velocity of the [\ion{Ne}{2}] line are mapped.  In H1, the intensity distributions of the 
Paschen-$\alpha$ (1.87 $\mu$m) and [\ion{Ne}{2}] lines are significantly different, which suggests 
a strong variation of extinction across the \ion{H}{2} region of $A_K$$\sim$0.56. 	
The radial velocity distributions across these \ion{H}{2} 
regions are consistent with the predictions of a bow-shock model for H1 
and the pressure-driven 
model for H2.  Furthermore, we find a concentration of bright 
stars in H2. These stars have similar $H$-$K_s$ 
colors and can be explained as part of a 2 Myr old stellar cluster. 
H2 also falls on the orbit of the molecular clouds, suggested to be around Sgr A$^*$.  
Our new results confirm what we had previously suggested:  
the O supergiant P114 in H1 is a runaway star, moving towards us 
through the -30$-$0 {km s$^{-1}$} molecular cloud, whereas the O If star P35 in H2 
formed {\sl in-situ}, and may mark the position of a so-far unknown small star cluster formed within
the central 30 pc of the Galaxy. 
\end{abstract}

\section{Introduction}
The Galactic Centre (GC) is the closest galactic nucleus 
($\sim$8.0 kpc,~\citealt{ghe08,gil09,gen10}), $\sim$100 times closer than the nearest 
nucleus of another spiral galaxy, the Andromeda galaxy. 
The GC therefore provides us with a unparalleled template for studying  
remote galactic nuclei.  In addition, understanding star formation within the GC helps in studies of
high-redshift galaxies, because their ISM conditions are likely similar~\citep{kru13}. 

The collection of giant molecular clouds located within a radius of $\lesssim$200 pc from the black hole Sgr A$^*$ are 
known as 
`Central Molecular Zone' (CMZ).  This region contains  
a few times $10^7$ M$_{\odot}$ of molecular hydrogen, mostly 
concentrated in some of the densest clouds in the galaxy~\citep{mor96,lau02,fer07}.  Even with this huge 
amount of gas, the star formation rate in the CMZ is 
an order of magnitude lower than that expected from 
empirical scaling relations~\citep{lon13}.   The low star formation rate in the GC is thought to be related to the observed high turbulent 
pressure~\citep{kru14}. 

Several dynamical models have been developed to explain the spatial and radial velocity 
distributions of molecular clouds in the GC and how they could collapse to form stars in this 
extreme environment.~\citet{mol11} find 
that molecular clouds in the GC appear to be aligned along an elliptical, twisted ring. They propose that these clouds 
could follow a closed elliptical orbit.~\citet{kru15} propose a different model, with open orbits, which explains  
why along some lines-of-sight there are three resolved velocity components. 
They also argue that their model can naturally explain the star formation stages 
from the `Brick' 
(no hot protocores,~\citealt[][and reference therein]{mil15}) to Sgr B2 (dense star formation regions). 
During the pericentre passage of Sgr A*, the central massive black hole, these dense clouds are compressed and 
star formation was triggered. They also claim that two of the three massive star clusters
found in the GC, the Arches and Quintuplet clusters~\citep[$\sim$2 and 4 Myr old, ][]{fig99,fig02}, 
are aligned with the proposed orbit of the molecular clouds. 

Observational evidence confirming this dynamical model is, however, still missing.  We 
have yet to find evidence of star clusters with ages between the Arches ($\sim$2-5 Myr) and Sgr B2 ($\lesssim$1 Myr), as would be 
expected based on the model of~\citet{kru15}, and assuming that  
the molecular clouds in the CMZ continuously pass by Sgr A*. The lack of adequate observational 
data is mostly due to the fact that the GC is obscured in the ultraviolet and optical bands (A$_{\rm V}\gtrsim$30) and the near-infrared (NIR) color similarities of 
main sequence stars and evolved low-mass stars. 
We were able to overcome this problem in our \hst/NICMOS GC survey~\citep{wan10,don11}, and succeeded in identifying 100 
apparently isolated candidates 
of evolved massive stars 
from their intensity enhancement near 1.87 $\mu$m due to the 
Paschen-$\alpha$ and He II lines. Some of these stars are associated with nearby \ion{H}{2} 
regions and may possibly mark so-far unknown young star forming regions. 
In the follow-up \gemini\ GNIRS/NIFS study of a small sample of eight 
evolved massive stars near the Arches cluster,~\citet{don15} found
that P35, an O If star in the H2 \ion{H}{2} 
region, seems to follow the nearby -30-0 {\rm km s$^{-1}$} molecular cloud and 
could have formed {\sl in-situ}. This suggestion was supported 
by the radial velocity pattern of the ionized gas along the long-slit of \gemini\ GNIRS.  However, 
most of an additional seven nearby stars seem to be interlopers that do not appear to be dynamically related to their associated 
\ion{H}{2} regions.  For example, 
the dynamical pattern of the ionized gas in the nearby H1 \ion{H}{2} region (55\arcsec , i.e. 2.2 pc, 
away from H2) can be explained by a bow-shock model, which indicates that P114, inside 
H1, is a runaway star. 

In this paper, we complement our previous \gemini\ GNIRS long-slit spectra with 
new Texas Echelon Cross Echelle Spectrograh~\citep[TEXES,][]{lac02} 
observations to map the 2-D distribution of the 
radial velocities of the ionized gas in the H1/H2 regions. We use these data to further study 
the dynamics of the ionized gas in these two \ion{H}{2} regions 
and how they impact on our understanding of the star formation 
processes in the CMZ. Our data are described in \S\ref{s:data} and the results 
are presented in \S\ref{s:result}. We discuss the physical 
origins of these two \ion{H}{2} regions, a new potential star cluster, and how our new results 
fit into the current concept of the orbit of molecular clouds in the GC in \S\ref{s:discussion}. 
A summary of the work is given in \S\ref{s:summary}. 

\section{Observations and Data Reduction}\label{s:data}
\subsection{\irtf/TEXES observations}
We observed the H1 and H2 regions with 
TEXES  
on the NASA 3.0 m Infrared Telescope Facility (\irtf )\footnote{Infrared Telescope Facility 
is operated by the University of Hawaii under contract NNH14CK55B with the National 
Aeronautics and Space Administration.} 
on Mauna Kea in April, 2015. TEXES is a high-resolution 
mid-infrared (5-25 $\mu$m) 
spectrograph. During our observations, TEXES was centered on 
the [\ion{Ne}{2}] (12.8 $\mu$m) line. Compared to previously used 
hydrogen recombination lines, [\ion{Ne}{2}] suffers less 
from foreground absorption (for example, $A_{12.8\mu m}\sim$1.34, 
$A_{2.166\mu m}\sim$2.49, $A_{1.875\mu m}\sim$3.55 toward the GC, 
from~\citealt{fri11}). Another advantage of using the [\ion{Ne}{2}] line 
is that for gas with $\sim$10$^4$ $K$ temperature, 
velocity broadening introduced by thermal motion 
in the hydrogen recombination lines is $\sim$21.4 {km s$^{-1}$}, 
but only 4.8 {\rm km s$^{-1}$} for the Ne$^+$ ions, enabling us to more 
accurately determine the radial velocities of the ionized gas~\citep{zhu05}.

\begin{figure*}[!thb]
  \centerline{    
       \epsfig{figure=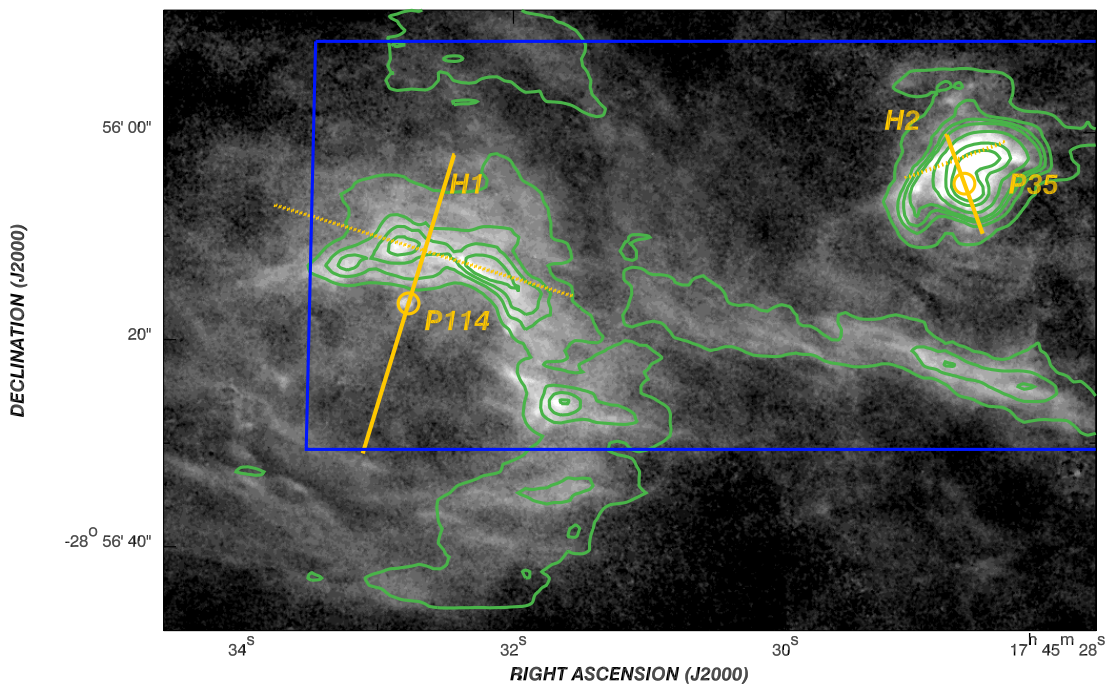,width=1.0\textwidth,angle=0}
       }
 \caption{The contour lines outline the intensity of the [\ion{Ne}{2}] (12.8 $\mu$m) line overlaid on the 
 \hst/NICMOS Paschen-$\alpha$ image of the H1 and H2 regions. 
 The contour levels are at 5\%, 8\%, 
 10\%, 12\%, 20\%, 40\% and 60\% of the peak intensity, 
 0.67 $erg$ $s$$^{-1}$ $cm$$^{-2}$ $sr$$^{-1}$. The blue box outlines the area 
 mapped in the high-resolution mode. The yellow circles mark 
 the two evolved massive stars, P114 and P35, associated with
  the H1/H2 \ion{H}{2} regions~\citep{don15}. 
 The yellow solid lines represent our proposed (approximately) symmetry 
 axes of these two \ion{H}{2} regions, while the dashed 
 lines show the perpendicular directions. The intensity distributions along these lines  
 are presented in Fig.~\ref{f:inten_cut}.}
\label{f:2d_image}
 \end{figure*}

For our observations, 
TEXES was operated in high-resolution and mid-resolution 
modes with resolutions of $\sim$4 {\rm km s$^{-1}$} and 
$\sim$25 {\rm km s$^{-1}$} along 
1.4\arcsec$\times$7.5\arcsec\ and 1.4\arcsec$\times$55\arcsec\ slits, respectively. 
The mid-resolution data completely covers both  
\ion{H}{2} regions, while the high-resolution mode observations missed the southeast 
corner of H1 (see Fig.~\ref{f:2d_image}). The two different resolutions enable us to 
to study the distributions of the intensity and radial velocity , respectively. 
We scanned the two \ion{H}{2} regions
using a north-south oriented slit 
from west to east with a spatial step size of 0.7\arcsec . 
The integration at each pointing is approximately 
15s.  An additional 10 pointings before the scientific observations
were used to measure the sky background. 
The spatial resolution was determined by seeing 
and for these observations is $\sim$1.4\arcsec .
 
We reduced the data with the standard TEXES pipeline 
procedure~\citep{lac02}. First, we corrected 
optical distortions, performed flat fielding, applied bad pixel 
masking, and then removed the cosmic-ray spikes. Second, through 
measurements of an ambient temperature blackbody before each set 
of observations, we derived radiometric calibration, which was used to 
translate the image units from {\rm counts} to  {\rm ergs cm$^{-2}$ s$^{-1}$ sr$^{-1}$ (cm$^{-1}$)$^{-1}$}. 
The uncertainty of the measured intensity is mostly systematic and roughly 20\%. 
Third, we used the sky emission lines to calibrate the wavelength, 
with an accuracy of $\sim$1 {\rm km s$^{-1}$}. Fourth, the sky background 
was subtracted from the observation of each pointing. Finally, multiple pointings 
were cross-correlated and added to produce the final data cube. 

\subsection{\hst/NICMOS GC Paschen-$\alpha$ Image}
The Paschen-$\alpha$ line (1.87 $\mu$m) image 
of the \ion{H}{2} regions H1 and H2 
is from our survey of the GC with the 
Near-Infrared Camera and Multi-Object Spectrometer (NICMOS) of the 
Hubble Space Telescope (\hst ). This survey~\citep{wan10} mapped the central 
36\arcmin$\times$15\arcmin\ of the GC with two narrow-band filters, 
F187N and F190N. The former band includes both the Paschen-$\alpha$ emission from the 
\ion{H}{2} regions and the stellar continuum at 1.87 $\mu$m, while the latter band just includes 
the nearby continuum at 1.90 $\mu$m. The pixel size of the final 
mosaic at these two bands is 0.1\arcsec/pixel with a spatial resolution 
of 0.2\arcsec . Source detection has been performed for individual pointings 
and filters with {\small`Starfinder'} software~\citep{dio00}, which 
is then merged into a master catalog. We scaled the F190N 
mosaic, which was subtracted from the F187N mosaic to get the pure 
Paschen-$\alpha$ emission. The detailed 
descriptions of the survey and the data analysis procedures are given in~\citet{don11}.  

\subsection{Broad-band photometry from \vlt/HAWKI survey of the GC}
\vlt/HAWKI survey of the GC is a large 
ESO program (Program ID 195.B-0283, PI, Rainer Sch{\"o}del) and maps the central 
46\arcmin$\times$14\arcmin\ field, plus several 8\arcmin$\times$5\arcmin\ fields 
along and perpendicular to the Galactic Plane, with the $J$, $H$ 
and $K_s$ filters, using the \vlt/HAWKI camera. Its main feature is that this 
survey uses 
holographic imaging technique~\citep{sch13}, which overcomes the seeing and 
makes the observations reach a spatial resolution of 0.2\arcsec , 
similar to our \hst/NICMOS survey. 
Due to the broad-band filters used and a bigger collecting area,
 this \vlt\ survey is deeper than the \hst\ one. 
The detailed descriptions of the survey and 
data reduction procedures are given in Nogueras-Lara et al. (2017, in preparation). For the bright 
sources, such as P35 and P114, which saturate in the \vlt/HAWKI survey, we use the photometry from 
the SIRIUS survey\footnote{Simultaneous 
3-color InfraRed Imager for Unbiased Surveys (SIRIUS) was taken by the Infrared Survey Facility (IRSF) 
in South Africa, with a pixel scale of 0.45\arcsec~\citep{nag03}. 
The survey includes the region $|l|$ $<$ 2 degree and $|b|$ $<$ 1
degree, with an angular resolution $\sim$ 1.2\arcsec\ in the $J$ band, 
better than that of 2MASS, $\sim$2\arcsec.}, which has similar filter transmission curves. 

\section{Results}\label{s:result}
\subsection{Integrated Intensity}
Fig~\ref{f:2d_image} shows the Paschen-$\alpha$ image with the contours of 
integrated [\ion{Ne}{2}] line intensity overlaid. The H1 and H2 regions are clearly seen with a size of 
$\sim$ 40\arcsec\ (1.6 pc) and 15\arcsec\ (0.6 pc). The overall structures of the spatial 
distributions of Paschen-$\alpha$ and [\ion{Ne}{2}] line intensities are very similar. 
H1 shows a cometary structure and is limb-brightened. The peak value 
of [\ion{Ne}{2}] in H1 is 0.07 {\rm ergs cm$^{-2}$ s$^{-1}$ sr$^{-1}$}. The 
only evolved massive star identified 
from our previous \hst/NICMOS GC survey~\citep{don12}, P114, an O supergiant, 
is not located in the center of the cavity. From the projected distance (5\arcsec .5, i.e. 0.21 pc),  
P114 is closer to the northern rim of the ionized filaments, which is 
also the brightest part of this \ion{H}{2} region. In contrast, the southwest rim is substantially fainter. 
The northern rim consists of two parts: the eastern and the western parts; the latter 
is brighter. On the 
other hand, H2 represents a fan-like structure. The brightest part of H2 is very 
close to the central star P35, with a peak value of 0.67 {\rm ergs cm$^{-2}$ s$^{-1}$ 
sr$^{-1}$}. 
At the northeastern part the intensity sharply decreases as we move away from the star,  
while at the southwestern part the intensity decreases slowly.  From the morphology of H1 and H2, 
we roughly infer a -17 and 20 degree position angles from north to east for the symmetry axes 
of these two \ion{H}{2} regions. 

\begin{figure*}[!thb]
  \centerline{    
       \epsfig{figure=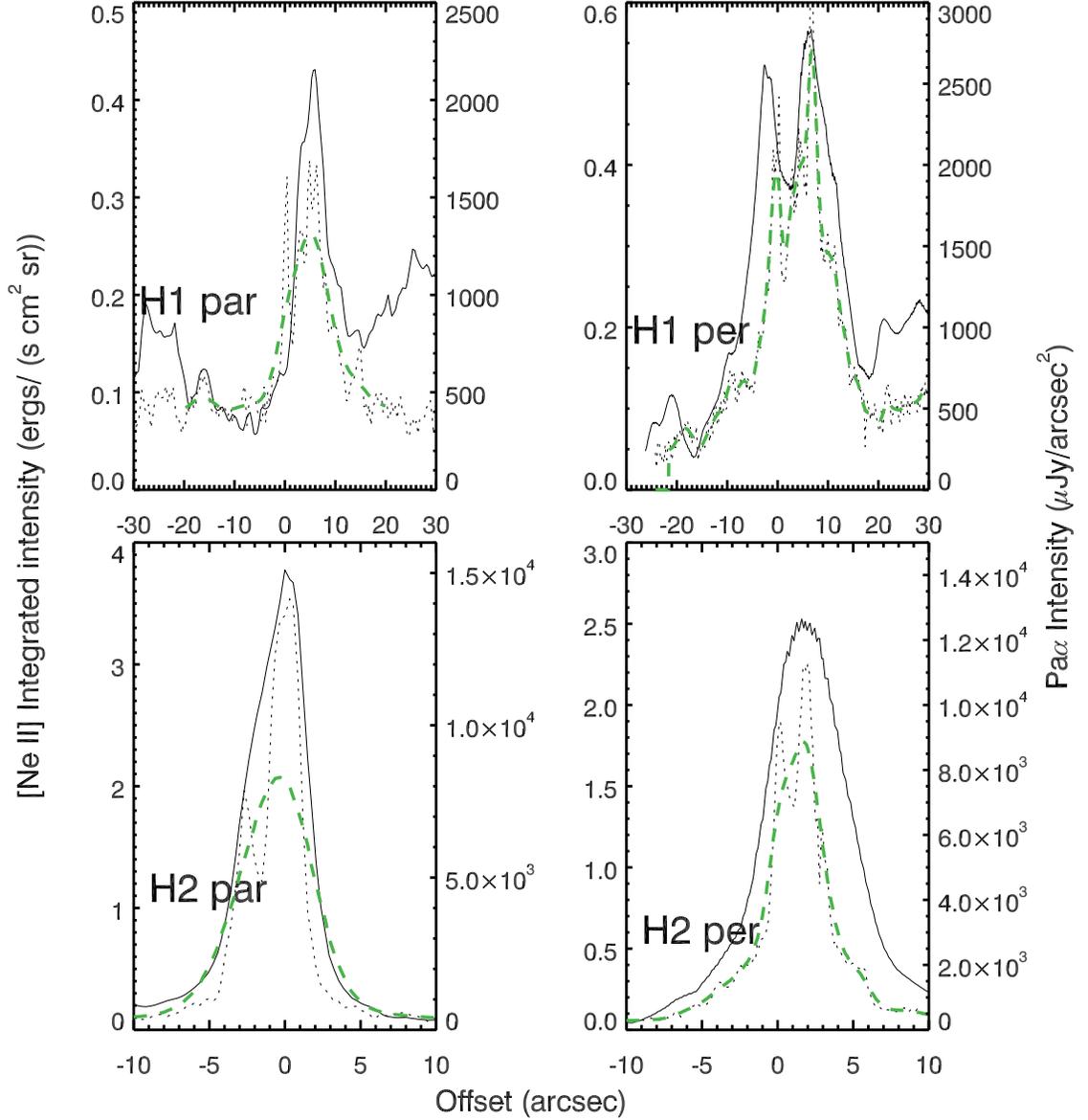,width=1.0\textwidth,angle=0}
       }
 \caption{Integrated intensity distributions along the yellow solid (left panels) 
 and dashed (right panels) lines in Fig~\ref{f:2d_image} of the H1 (top panels) and H2 (bottom panels). 
 The zero offsets in the left panels are at P114 and P35, respectively, 
 while those in the right panels are at the interactions of the yellow solid and dashed lines in Fig~\ref{f:2d_image}. 
 The solid curves are for the [\ion{Ne}{2}] line, while the dotted curves are for the Paschen-$\alpha$ line. 
 The green dashed curves represent the intensity distributions of Paschen-$\alpha$ line convolved with a 
 Gaussian kernel of 1.4\arcsec\ to match the resolution of the [NeII] observations.}
\label{f:inten_cut}
 \end{figure*}

\begin{figure*}[!thb]
  \centerline{    
       \epsfig{figure=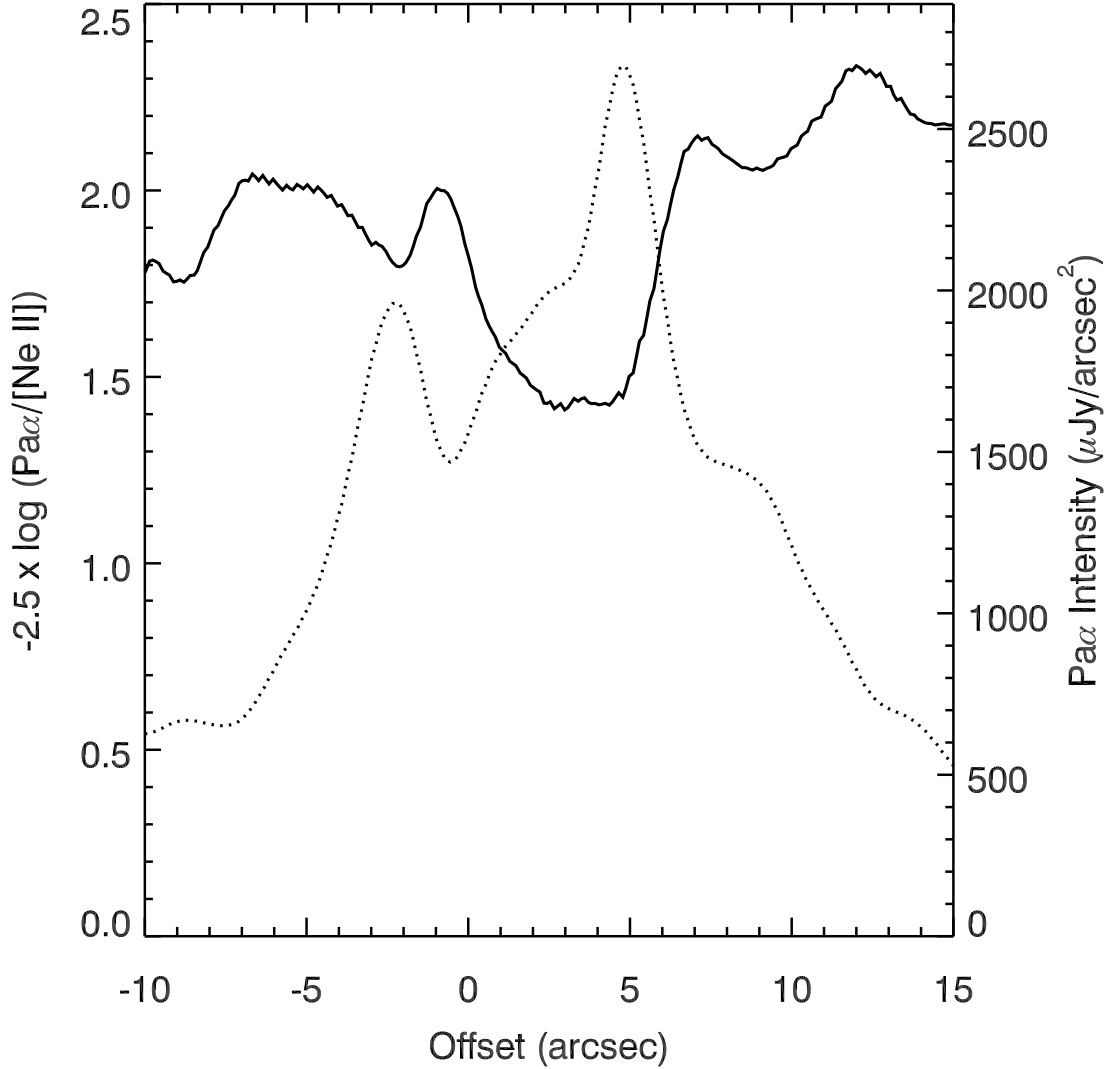,width=1.0\textwidth,angle=0}
       }
 \caption{The intensity ratio (solid line) of Paschen-$\alpha$ line (in units of 2$\times10^4$ $\mu$Jy/arcsec$^2$) to [\ion{Ne}{2}] 
 line (in units of $erg$ $s$$^{-1}$ $cm$$^{-2}$ $sr$$^{-1}$) as a function of the offset along 
 the cut perpendicular to the symmetry axis of the H1 (the yellow dashed line on the left of Fig~\ref{f:2d_image}). 
 The dashed line 
 is the integrated intensity distribution of the Paschen-$\alpha$ line, convolved with a Gaussian 
 kernel of 1.4\arcsec . The positive offset is toward the southwest. }
\label{f:extin}
 \end{figure*}

Fig.~\ref{f:inten_cut} shows the intensity distributions along several cuts parallel or 
perpendicular to the symmetry axes of H1 and H2 for both [\ion{Ne}{2}] (black solid lines) 
and Paschen-$\alpha$ (black dotted lines). Considering that the angular resolution of the 
Paschen-$\alpha$ image is much better than that of the [\ion{Ne}{2}] line image (see \S\ref{s:data}), we 
convolve the intensity distributions of the Paschen-$\alpha$ image by a Gaussian kernel with a width of 
1.4\arcsec . For all the cuts shown in Fig.~\ref{f:inten_cut}, the peaks of the 
intensity distribution of the [\ion{Ne}{2}] and smoothed Paschen-$\alpha$ match very well. In 
the cut perpendicular to the symmetry axis of H1, we see the intensity distribution has two 
distinctive peaks at both wavelengths; with the western peak at 
an offset of $\sim$5\arcsec\ stronger than that at $\sim$-3\arcsec. 
However, the intensity ratio between these two peaks seems to be smaller 
in the [\ion{Ne}{2}] observation. Even if we 
correct for the spatial resolution differences between the Paschen-$\alpha$ and 
[\ion{Ne}{2}] observations, the difference between these two datasets is still very significant. In order to 
demonstrate this clearly, in Fig.~\ref{f:extin}, we show the intensity ratio between the Paschen-$\alpha$ 
and [\ion{Ne}{2}] lines, which as can be seen varies significantly along the cut. The ratio is smallest at offset $\sim$3\arcsec\ to 5\arcsec\ 
and is roughly constant away from the brightest part. 
Extinction, gas excitation and abundance could cause this variation 
of the intensity ratio.  We do not expect any differences in the 
Ne abundance in this small physical region ($<$0.2 pc). Meanwhile, 
the eastern part (offset $\sim$0\arcsec ) of the northern rim is 
closer to P114 than the western part (offset $\sim$5\arcsec ). 
We also do not find any emission line stars near the western part. 
Therefore, we do not believe that the small 
ratio at the western part is caused by higher gas excitation. We suggest 
that near the brightest 
Paschen-$\alpha$ peak, the 
foreground extinction is smaller than that in the neighboring 
region by $A_{F190N}\sim$0.8 mag, i.e. $A_K\sim$0.56 mag and $A_{12.8\mu m}$=0.3 mag, 
assuming the extinction curve given in~\citet{fri11}.  

\begin{figure*}[!thb]
  \centerline{   
       \epsfig{figure=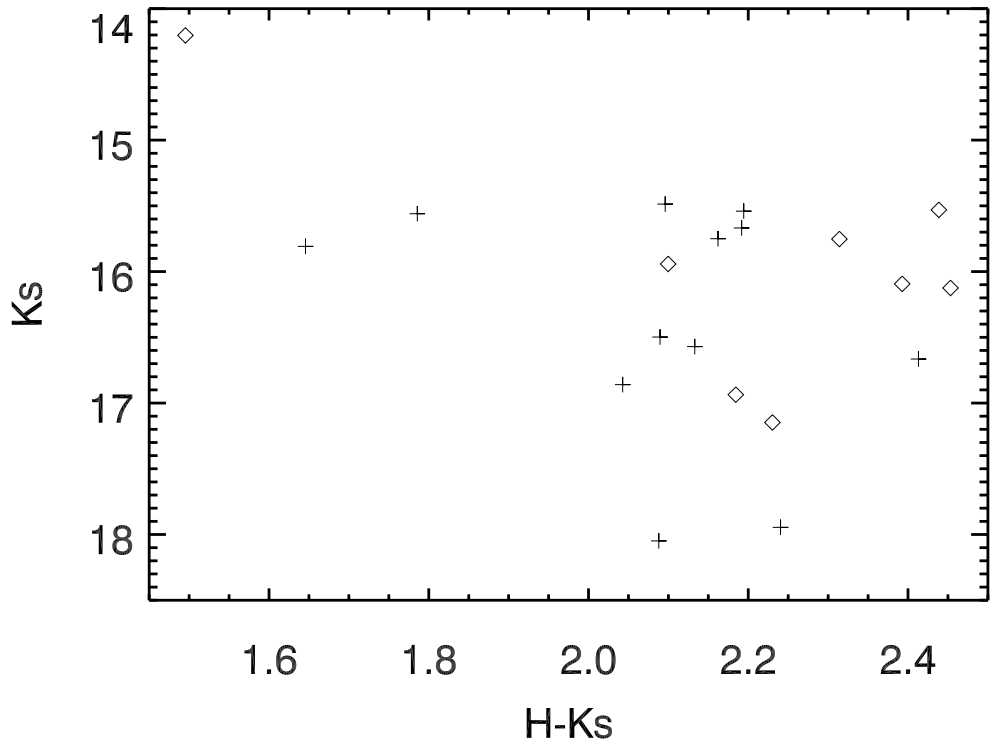,width=1.0\textwidth,angle=0}
       }
 \caption{In this CMD ($H$-$K_s$ vs $K_s$), 
 the plus and diamond symbols represent the stars detected from the 
 \vlt/HAWKI survey within 2\arcsec\ of the western and eastern parts of the 
 northern rim in H1.}
 \label{f:H1_extin_color}
 \end{figure*}

The different extinctions at the two parts are 
also supported by the colors of nearby stars. 
Fig.\ref{f:H1_extin_color} presents the color magnitude 
diagram (CMD, $H$-$K_s$ vs $K_s$) of stars found from the \vlt/HAWKI 
survey within 2\arcsec\ of the two parts. 10 and 7 out of 18 and 16 detected sources 
have both $H$ and 
$K_s$ magnitudes, as well as $H$-$K_s$$>$2, indicating that they are in the GC. 
The colors of the stars in the western part (`pluses') are tighter and bluer 
than those in the eastern part (`diamonds'). The difference of the median colors ($H$-$K_s$) 
of the stars in these two parts is 0.15 mag, i.e. $A_K\sim0.21$ mag. The 
reason why this value is smaller than that derived from the intensity ratio of 
Paschen-$\alpha$ 
and [\ion{Ne}{2}] lines could be because some stars with higher extinctions are 
not found in the \vlt/HAWKI survey due to the detection limit.

\subsection{Velocity Channel Map}
\begin{figure*}[!thb]
  \centerline{    
       \epsfig{figure=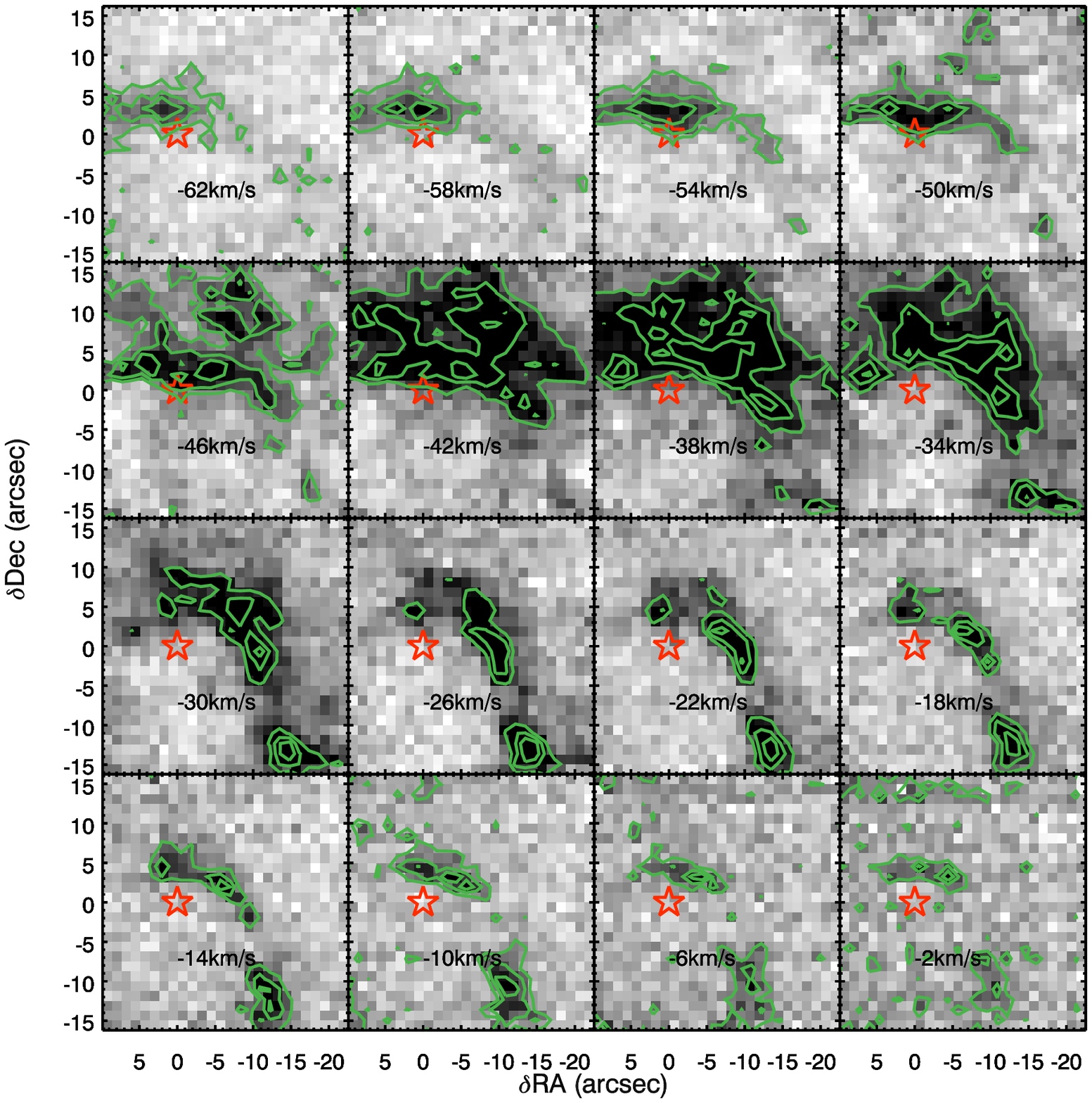,width=1.0\textwidth,angle=0}
       }
 \caption{Channel maps of the [\ion{Ne}{2}] line surface brightness of H1, with the coordinates relative to P114. The velocity is relative to the local standard of rest. 
 The velocity width of each map  
 is 4 {\rm km s$^{-1}$}. Green contour levels are at 80\%, 60\% and 40\% of the peak value of the surface brightness, i.e. 0.59 erg s$^{-1}$ cm$^{-2}$ sr$^{-1}$ (cm$^{-1}$)$^{-1}$.}
 \label{f:h1_vel}
 \end{figure*}

\begin{figure*}[!thb]
  \centerline{    
       \epsfig{figure=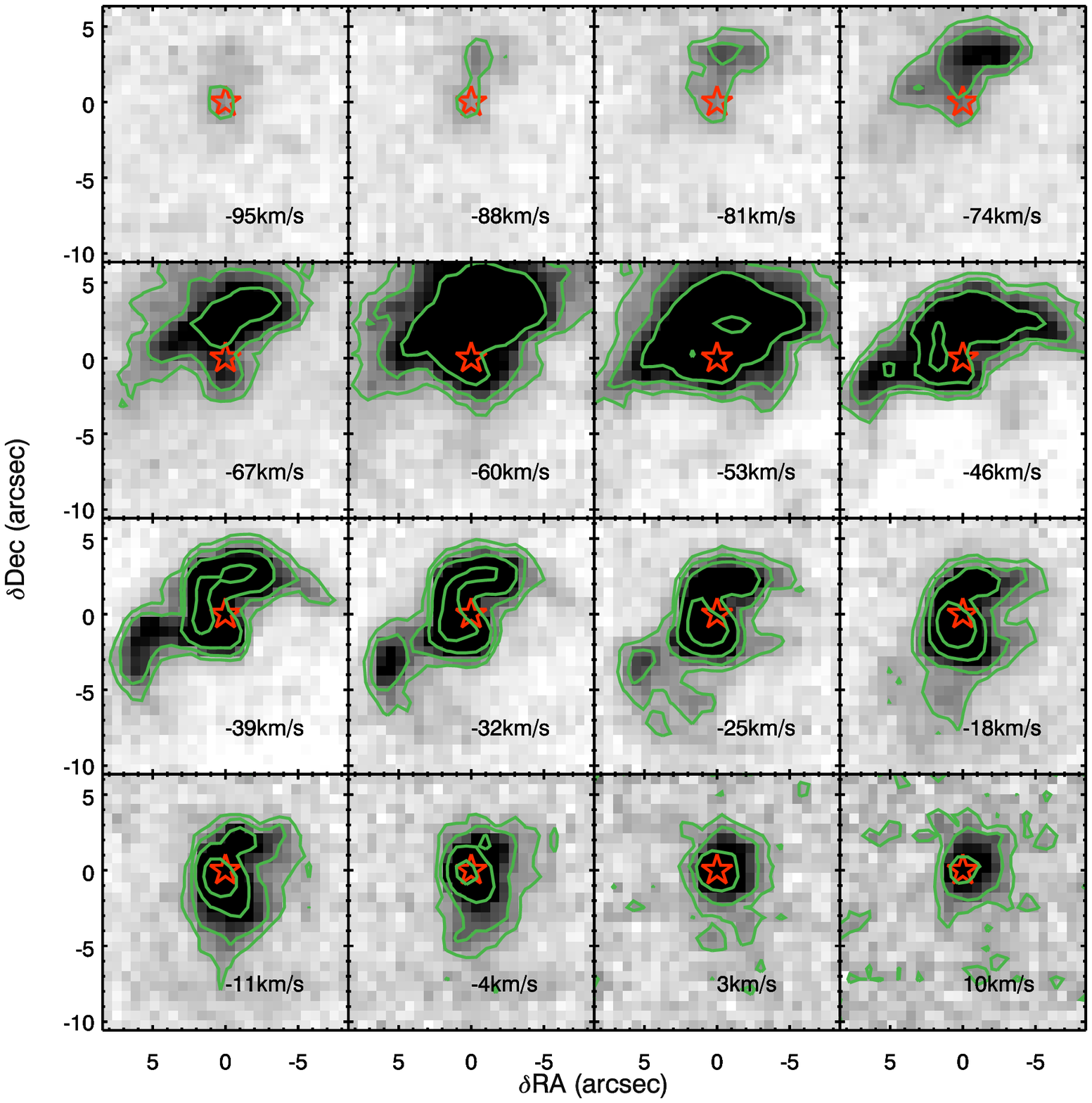,width=1.0\textwidth,angle=0}
       }
 \caption{Channel maps of the [\ion{Ne}{2}] line surface brightness of H2, with the coordinates relative to P35. The velocity is relative to the local standard of rest. 
 The velocity width of each map 
 is 7 {\rm km s$^{-1}$}. Green contour levels are at 30\%, 10\%, 5\% ad 3\% of the peak value of the surface brightness, i.e. 3.66 erg s$^{-1}$ cm$^{-2}$ sr$^{-1}$ (cm$^{-1}$)$^{-1}$.}
  \label{f:h2_vel}
 \end{figure*}

Figs~\ref{f:h1_vel} and~\ref{f:h2_vel} present the velocity channel 
maps of H1 and H2, respectively, derived from the high-resolution observations. 
In H1, most of the high velocity \ion{Ne}{2} 
components (-62 {\rm km s$^{-1}$} to -50 {\rm km s$^{-1}$}) concentrate 
on the northern part, close to the central star, whereas the intermediate velocity gas 
(-50 to -30 {\rm km s$^{-1}$}) extends to the northwestern part and the low velocity gas 
(-30 to -10 {\rm km s$^{-1}$}) along the ridge, but mostly in 
the southwestern 
part. The lower left portion (southeast) of the field covered by the observation completely 
lacks ionized gas emission. Unlike H1, H2 is bright along 
the line-of-sight toward P35 over the entire velocity range, indicating that the star is 
embedded in the ionized gas and its strong stellar wind is pushing surrounding  
gas away in all directions. The majority of the high velocity gas, 
is found in the northern part of the \ion{H}{2} region.  There does not appear 
to be a consistent gradient of 
higher to lower velocity gas that is correlated with position. 

\begin{figure*}[!thb]
  \centerline{    
       \epsfig{figure=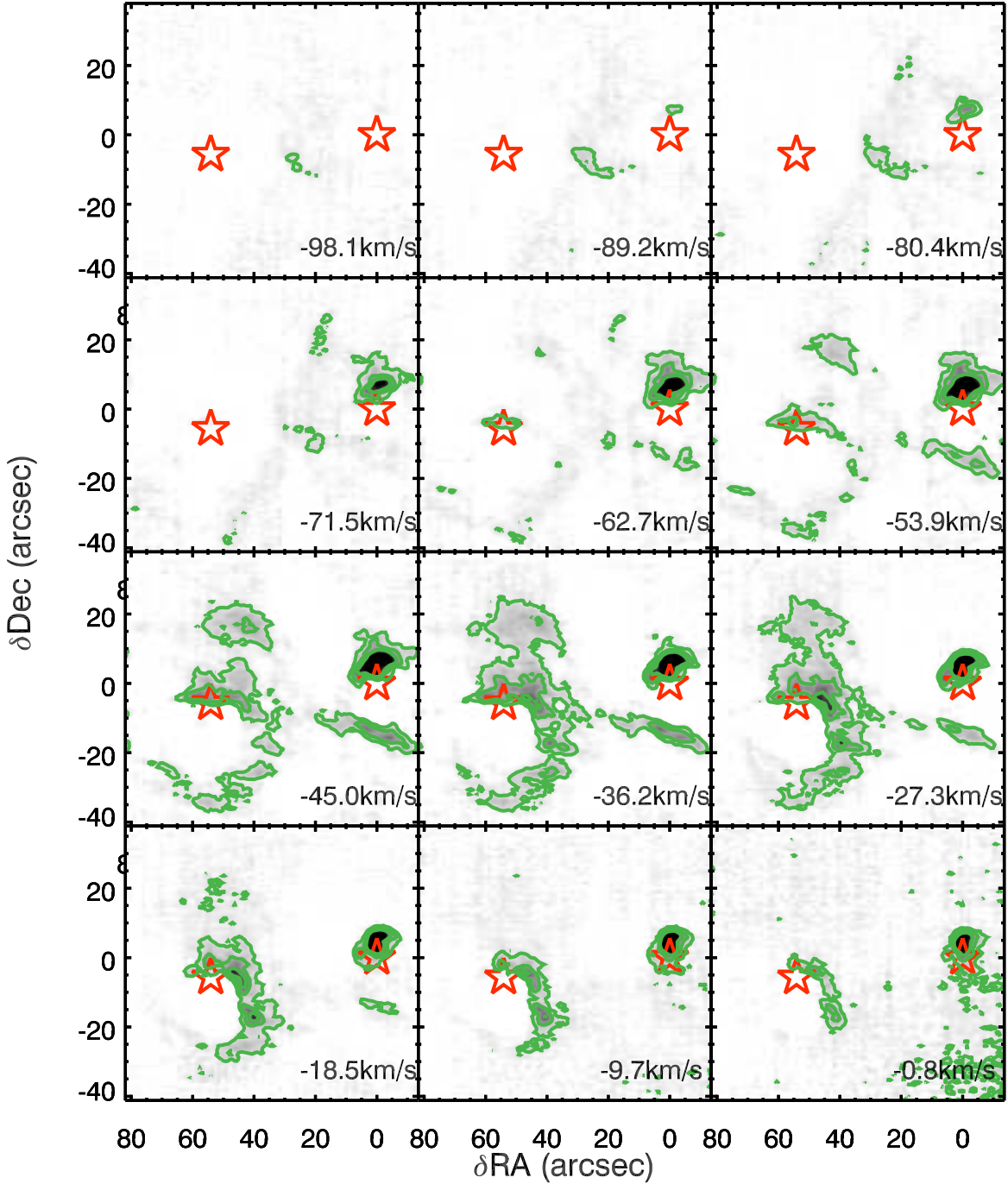,width=1.0\textwidth,angle=0}
       }
 \caption{Channel maps of the [\ion{Ne}{2}] line surface brightness derived from the med-resolution observation. 
  with the coordinates relative to P35. The field-of-view of 
 these images is the same as that  
 of Fig.~\ref{f:2d_image}. The velocity is relative to the local standard of rest. 
 The velocity width of each map 
 is 8.84 {\rm km s$^{-1}$}. Green contour levels are at 10\%, 5\% and 2\% of the peak value of the surface brightness, i.e. 1.76 erg s$^{-1}$ cm$^{-2}$ sr$^{-1}$ (cm$^{-1}$)$^{-1}$. The two red stars 
 represent the evolved massive stars, P114 and P35 embedded in H1 and H2, respectively. }
 \label{f:med_vel}
 \end{figure*}

In Fig.~\ref{f:med_vel}, we present the velocity channel maps derived from the 
mid-resolution observation. Its larger field-of-view and higher sensitivity 
provide us with more information about the relationship between the two \ion{H}{2} regions. 
Between H1 and H2, 
there is a weak high-velocity component ($\sim$-85 {\rm km s$^{-1}$}, 
($\delta$RA, $\delta$Dec)=(25\arcsec , -5\arcsec ) relative to P35). South of the H2 region observed in 
the high resolution observations, 
there is an ionized ridge with a velocity of $\sim$-40 {\rm km s$^{-1}$}). Similar ridges 
have also been seen in the well studied GC \ion{H}{2} region, Sgr A-A~\citep{mil11}. 
The structure of H1, especially the southern part, which is not 
covered by the high-resolution observation, is relatively bright at about the same 
velocity of $\sim$-40  {\rm km s$^{-1}$}. 

\subsection{Position-Velocity Diagram}
Position-velocity (PV) diagrams along the cuts parallel or perpendicular to 
the symmetry axes of H1 and H2 are presented in  
Figs~\ref{f:h1_par}$-$\ref{f:h2_pen}. 
In each figure, the cut in the first row passes through 
the central evolved massive star, P114 or P35.

The velocity field of H1 indicates the presence of an expanding gas 
bubble around P114. 
The PV diagrams along the cuts parallel to the 
symmetry axis (Fig.~\ref{f:h1_par}) show that the ionized gas 
far away from the central star (13\arcsec\ to 20\arcsec\ offset, northwest of P114) 
has -40 {\rm km s$^{-1}$}. Closer to P114, two 
dimmer components become apparent; one with a velocity of -50 {\rm km s$^{-1}$} and 
the other at $\geq$-20 {\rm km s$^{-1}$}. The two regions are offset  
$\sim$8\arcsec\ from the star along the cut, respectively. 
The -50 {\rm km s$^{-1}$} component appears only to the 
the east of the symmetry axis (the first and second panels of Fig.~\ref{f:h1_vel}). 
The $\geq$-20 {\rm km s$^{-1}$} component can be detected 
in all the three cuts, although it is slightly weaker along the symmetry 
axis. The PV diagrams along the cuts perpendicular to the 
symmetry axis (Fig.~\ref{f:h1_pen}) show that the high velocity 
component ($<$-40 {\rm km s$^{-1}$}) 
only exists 
on the eastern (left) part of the northern rim ($\sim$ 5\arcsec\ offset), with the large velocity dispersion 
is seen at 5\arcsec\ and -10\arcsec\ offsets; the radial velocity of the ionized gas,  
extending from -40 {\rm km s$^{-1}$} to -10 {\rm km s$^{-1}$}. We do not see a similarly coherent 
velocity structure at the -3\arcsec\ to 3\arcsec\ offsets. Therefore, the shell appears to be 
incomplete. 

In H2, the radial velocity 
of the ionized gas monotonically decreases from $\sim$-60 {\rm km s$^{-1}$} in the head of the \ion{H}{2} 
regions ($\sim$4\arcsec\ offset), to -10 {\rm km s$^{-1}$} around the central star along 
cuts parallel to the symmetry axis (Fig.~\ref{f:h2_par}). 
Along the cuts perpendicular to the symmetry axis, there is 
a large deviation of the radial velocity near the star. We also see 
a large negative velocity of $\sim$-60 {\rm km s$^{-1}$} located
$\sim$-3\arcsec\ to 5\arcsec\ from the star along the perpendicular cuts. 

\section{Discussion}\label{s:discussion}
We first compare our results with previous \gemini/GNIRS observations in \S\ref{ss:d_pre}. Then, 
we study the physical origins of H1/H2 in \S\ref{ss:d_h1h2}. Third, we 
discuss the implications of a new star cluster embedded in H2 in \S\ref{ss:d_c}. Finally,   
we discuss this new star cluster in the context of both the existing dynamical model of molecular 
clouds and the star formation in this extreme hostile galactic 
nuclear environment in \S\ref{ss:d_s}. 

\subsection{Comparison of the \irtf/TEXES and \gemini/GNIRS observations}\label{ss:d_pre}
\citet{don15} study the radial velocity of ionized gas along a \gemini/GNIRS 
long slit from the Br$\gamma$ line with a spectral resolution of  40 {\rm km s$^{-1}$} in H1 and H2. 
The spatial positions of the long slit are similar to the cuts parallel to the 
symmetric axes and passing the central stars given in 
Figs~\ref{f:h1_par} and~\ref{f:h2_par}. 
In H1, we found the radial velocity of the ionized gas 
is -39.6$\pm$2.7 {\rm km s$^{-1}$} at 
$\sim$6.6\arcsec\ and -72.7$\pm$2.9 {\rm km s$^{-1}$} at $\sim$2.7\arcsec . The former is 
consistent with the new PV diagram derived from the [\ion{Ne}{2}] line, while 
we do not detect the latter component in the TEXES dataset, mostly likely due to the low  
signal-to-noise ratio of our data at that velocity. 
Conversely, in the \gemini/GNIRS dataset, we do not resolve 
the dim -10 {\rm km s$^{-1}$} component seen in the first 
panel of Fig.~\ref{f:h1_par} due to poor 
spectral resolution.  In H2,~\citet{don15} find that the ionized gas peaks at -63.5$\pm$2.6 
{\rm km s$^{-1}$} at 2.9\arcsec\ away from the P35, decreasing to $\sim$0 {\rm km s$^{-1}$} near P35. This  
pattern is fully consistent with the one shown in the first panel of Fig.~\ref{f:h2_par}. 

\subsection{The origins of H1 and H2}\label{ss:d_h1h2}
Gas kinematics derived from the [\ion{Ne}{2}] line play a critical role in enabling us to constrain the 
underlying physical processes in these \ion{H}{2} regions. Different scenarios 
have been proposed to explain the structures of the \ion{H}{2} regions, such as bow shock and 
pressure driven scenarios. These scenarios could produce similar projected shapes of  
ionized gas~\citep{zhu08}. Consequently, the morphology alone is not enough to 
distinguish which of these scenarios is correct.
The two models, however, predict different radial velocities for ionized gas 
as related to nearby neutral gas.  

~\citet{don15} find that the `bow shock' and `pressure-driven' scenarios are able to explain 
the one-dimensional radial velocities of the ionized gas in H1 and H2, respectively. 
In the case of H1, a massive star is quickly moving into an existing
molecular cloud. Its strong stellar wind compresses the surface of the molecular cloud 
and produces a thin shell. In the case of H2, the star with the stellar wind
 is at rest relative to the nearby molecular cloud. This results in a fan-like
  \ion{H}{2} region due to the density gradient of the molecular cloud; the stellar wind
  breaks up the molecular cloud starting from the low density region and moves 
  towards us, while the bright part 
  of the \ion{H}{2} region represents higher densities. 
\citet{don15} reach the above conclusion from only 
  one-dimension measurements of the radial velocity of the ionized gas. Here, we  
  further test the scenarios with new \irtf/TEXES data. 
     
\begin{figure*}[!thb]
  \centerline{    
       \epsfig{figure=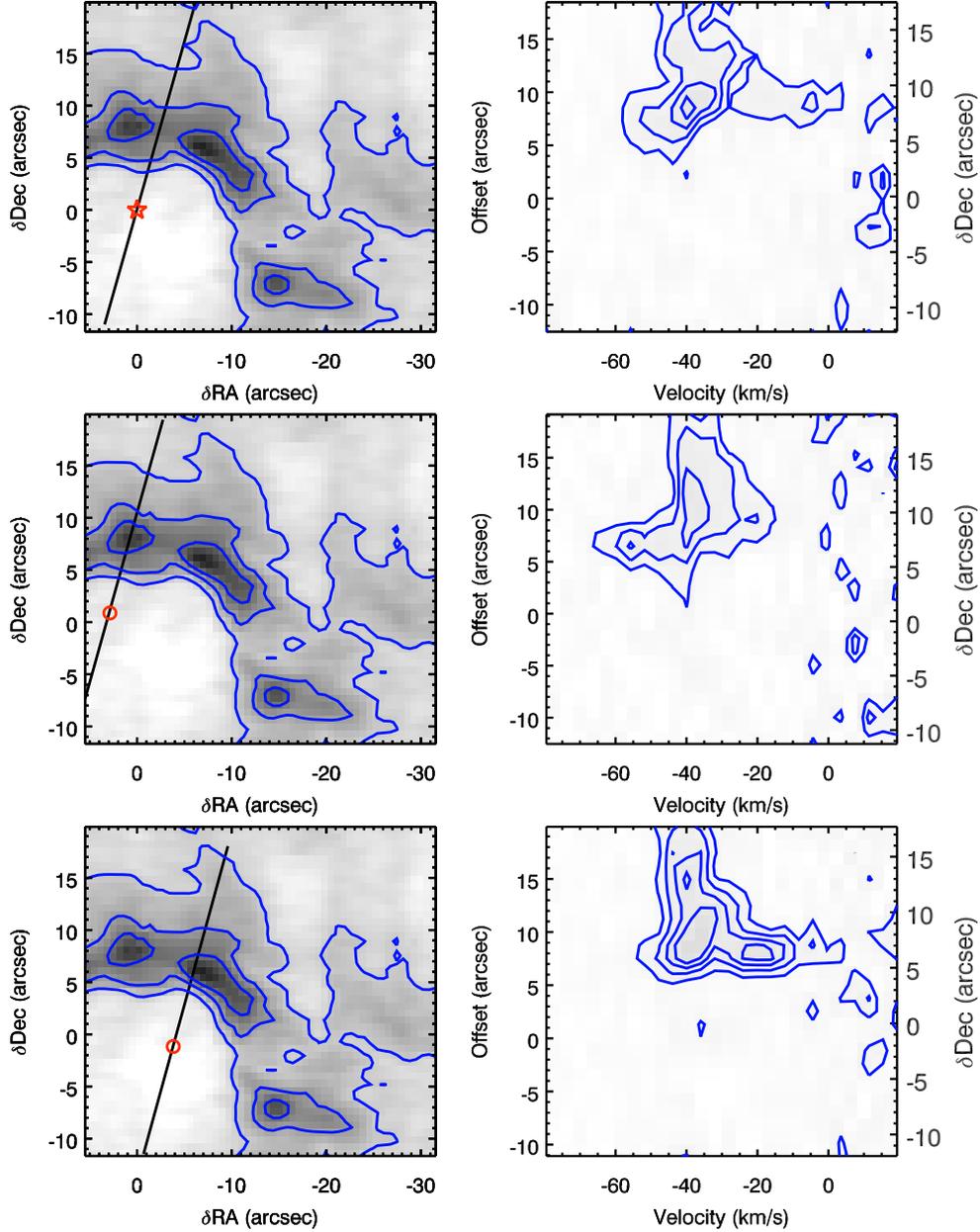,width=0.9\textwidth,angle=0}
       }
 \caption{Left: TEXES [\ion{Ne}{2}] integrated line flux map of H1. The black lines show the cut for 
 the PV diagram on the right. The star and circles are the zero offsets 
 along the cuts. The star also marks the location of P114. The locations along the 
 cut above (bellow) the star or circles have positive (negative) offsets. Blue contour 
 levels are at 70\%, 50\% and 30\% of the peak value of the integrated line flux map of H1, 
 0.1 {\rm ergs cm$^{-2}$ s$^{-1}$ sr$^{-1}$}. Right: 
 The PV diagram of the H1 along 
 the cuts parallel to the symmetry axis. Blue contours are drawn at 
 50\%, 40\%, 30\%, and 20\% of the peak value of the surface brightness of H1, i.e. 
 0.59 erg s$^{-1}$ cm$^{-2}$ sr$^{-1}$ (cm$^{-1}$)$^{-1}$. 
 In the first panel, the cut passes 
 P114, while in the second and third panels, the cuts are in 
 the 3\arcsec\ northeast (top left) and 
 4\arcsec\ southwest (low right) away from the cut in the first panel. The vertical axis is the offset 
 along the cut from the star or circles. In the right side of each panel, we show the corresponding $\delta$Dec given 
 in the left column.}
 \label{f:h1_par}
 \end{figure*}

\begin{figure*}[!thb]
  \centerline{   
       \epsfig{figure=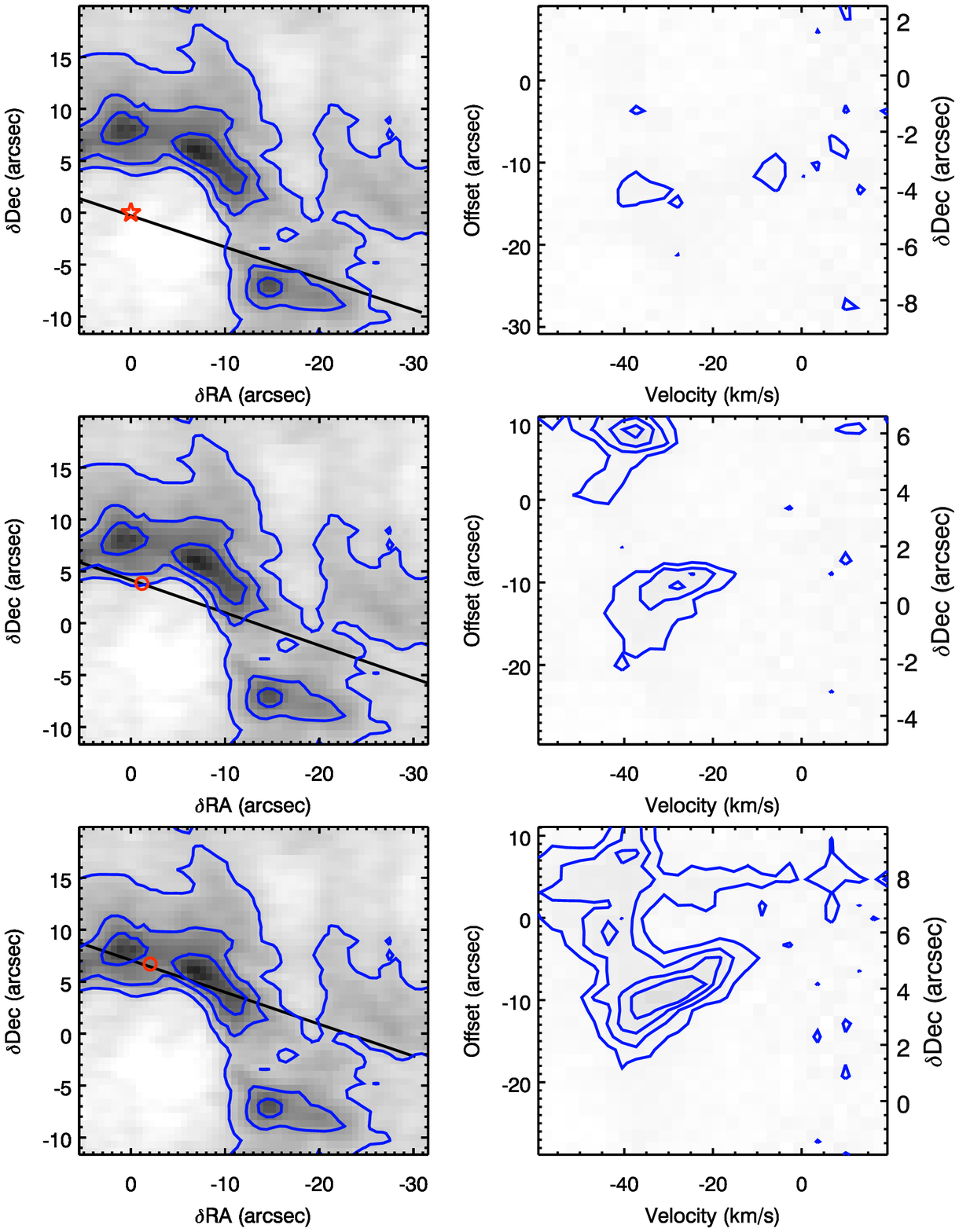,width=0.9\textwidth,angle=0}
       }
 \caption{The same as Fig.\ref{f:h1_par} except for the cut perpendicular to 
 the symmetry axis.  In the second and third panels, the cuts are in the 4\arcsec\ and 
 7\arcsec\ northwest (top right) away from the cut in the first panel. }
  \label{f:h1_pen}
 \end{figure*}

\begin{figure*}[!thb]
  \centerline{    
       \epsfig{figure=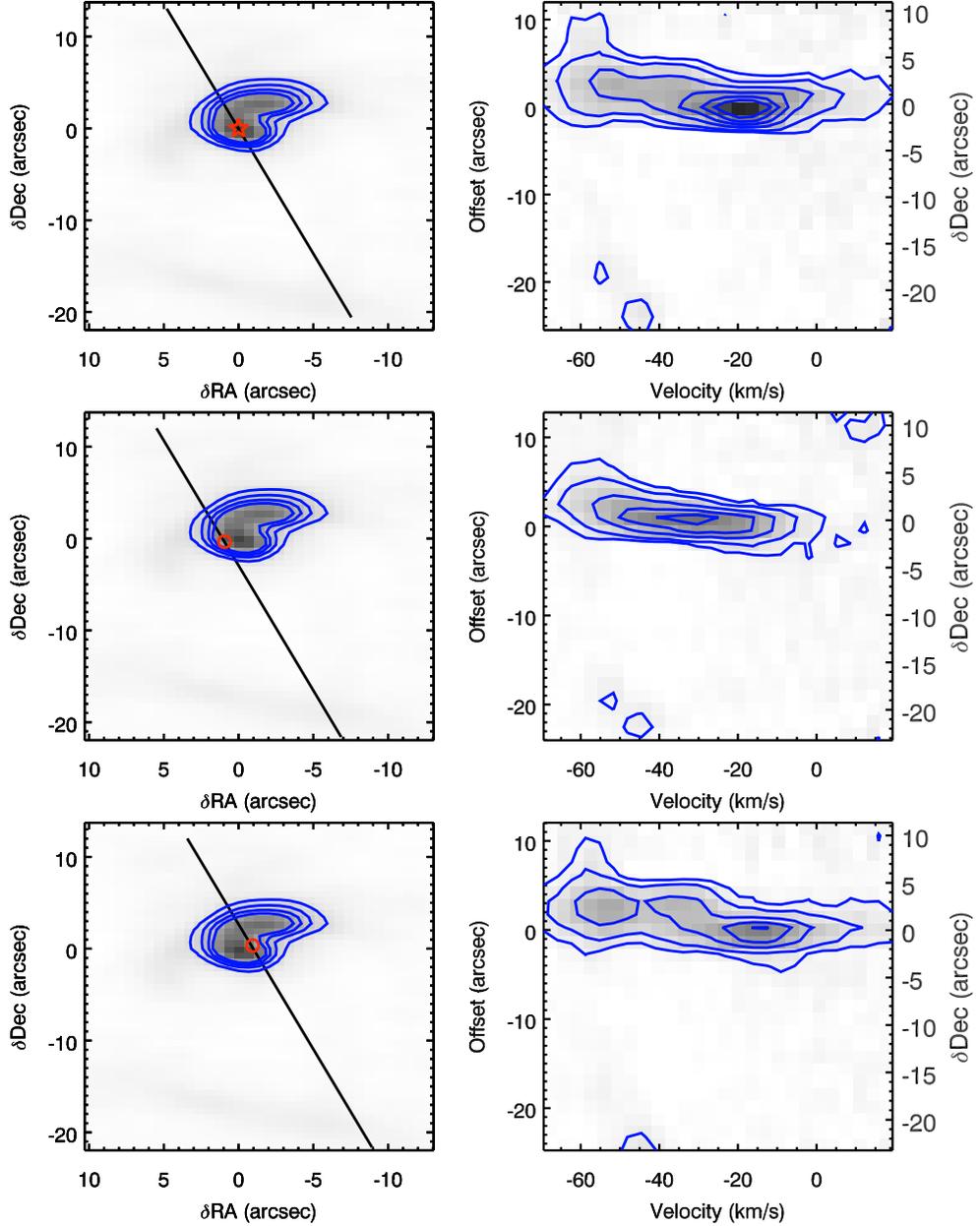,width=0.9\textwidth,angle=0}
       }
 \caption{The same as Fig.\ref{f:h1_par} except for H2. For the left panel, blue 
 contours are drawn at 
 50\%, 40\%, 30\%, and 20\% of the peak value of the integrated line flux map of H2, 
0.67 erg s$^{-1}$ cm$^{-2}$ sr$^{-1}$. For the right panel, blue contours are drawn at 
 50\%, 40\%, 30\%, 20\%, 10\% and 5\% of the peak value of the surface brightness of H2, i.e. 
 3.66 erg s$^{-1}$ cm$^{-2}$ sr$^{-1}$ (cm$^{-1}$)$^{-1}$. In the second and 
 third panels, the cuts are in the 1\arcsec\ southeast (low left) and 
 1\arcsec\ northwest (top right) away from the cut in the first panel.  }
  \label{f:h2_par}
 \end{figure*}

For H1,~\citet{don15} argue that P114 is behind the molecular cloud and is moving towards 
us with an angle of inclination between 0 and 90 degree. 
Each line-of-sight includes the emission from the front and back sides 
of the thin shell produced by the stellar wind. Ionized gas from the 
thin shell produced by the stellar wind is blueshifted relative to nearby neutral gas 
 ($\sim$-30 to -40 {\rm km s$^{-1}$}, according to Fig 9 in~\citealt{hen16}). 
This effect becomes increasingly 
apparent toward the central star. The  
velocity becomes more negative when moving closer to 
P114 in the PV diagram of Fig. 8 in~\citet{don15} (-72.7 {\rm km s$^{-1}$} component). 
A similar effect can be seen in the northern part of the H1 
region along the cuts parallel to the 
symmetry axis shown in 
Fig.~\ref{f:h1_par}. 
The -40 {\rm km s$^{-1}$} component behind the thin shell (15\arcsec\ offset in 
 Fig.~\ref{f:h1_par}) traces the 
velocity of the molecular cloud. 
When moving towards P114, the components with velocities smaller and bigger than -40 {\rm km s$^{-1}$} 
are from the front and back sides of the thin shell. In the latter case, the ionized gas is 
far and moves away from P114, although with a absolute velocity that is rather small. 
The same large range of the radial velocities is seen 
in the cut perpendicular to the symmetry axis and passing from the brightest 
part of the shell (the bottom panel of Fig.~\ref{f:h1_pen}), which should trace the 
emission from the two sides of the shell. The southern part of H1, which is far away from P114, could be the 
photo-ionized surface of 
the nearby molecular cloud, a result of the ultraviolet photons 
from the O supergiant P114. Therefore, the velocity derived from [\ion{Ne}{2}] line is similar to 
that of nearby molecular clouds. On the other hand, in a pressure-driven model, all the ionized 
gas moves towards us following the surface of the shell and no radial velocities bigger 
than -40 {\rm km s$^{-1}$} are expected in the PV diagram shown in Fig.~\ref{f:h1_par} 
and Fig.~\ref{f:h1_pen}.

One major concern in the bow shock interpretation of H1 in~\citet{don15} is 
correctly interpreting the 
foreground extinction toward P114 and its effect on the observed Paschen-$\alpha$ emission, 
since the scenario expects extra extinction from the foreground molecular cloud. 
One potential solution is that P114 has already reached the front side of the molecular cloud. This is supported by 
the fact that the western part of the northern rim in H1 shows $\sim$0.6 mag at $K_s$
(or $\sim$0.3 mag at 12.8 $\mu$m) less extinction than seen 
in the eastern part. 
If the molecular cloud in the western part of the 
northern rim was possibly already destroyed, this also explains why no large negative velocities are be 
seen there (offset -10-0\arcsec ), but are only found at $\sim$5\arcsec\ (see the last panel of Fig.~\ref{f:h1_pen}). 
Although the western part is further away from P114 than the eastern part in projected distance, 
the [\ion{Ne}{2}] line emission suffers from less foreground extinction by a factor of 1.3 in the western part, 
which explains why the brightest peak falls in the western part, but not the eastern part. 
Such a spatial variation of extinction is not expected by the pressure-driven model, in which we 
directly see the inner surface of molecular cloud facing the central star. 

\begin{figure*}[!thb]
  \centerline{   
       \epsfig{figure=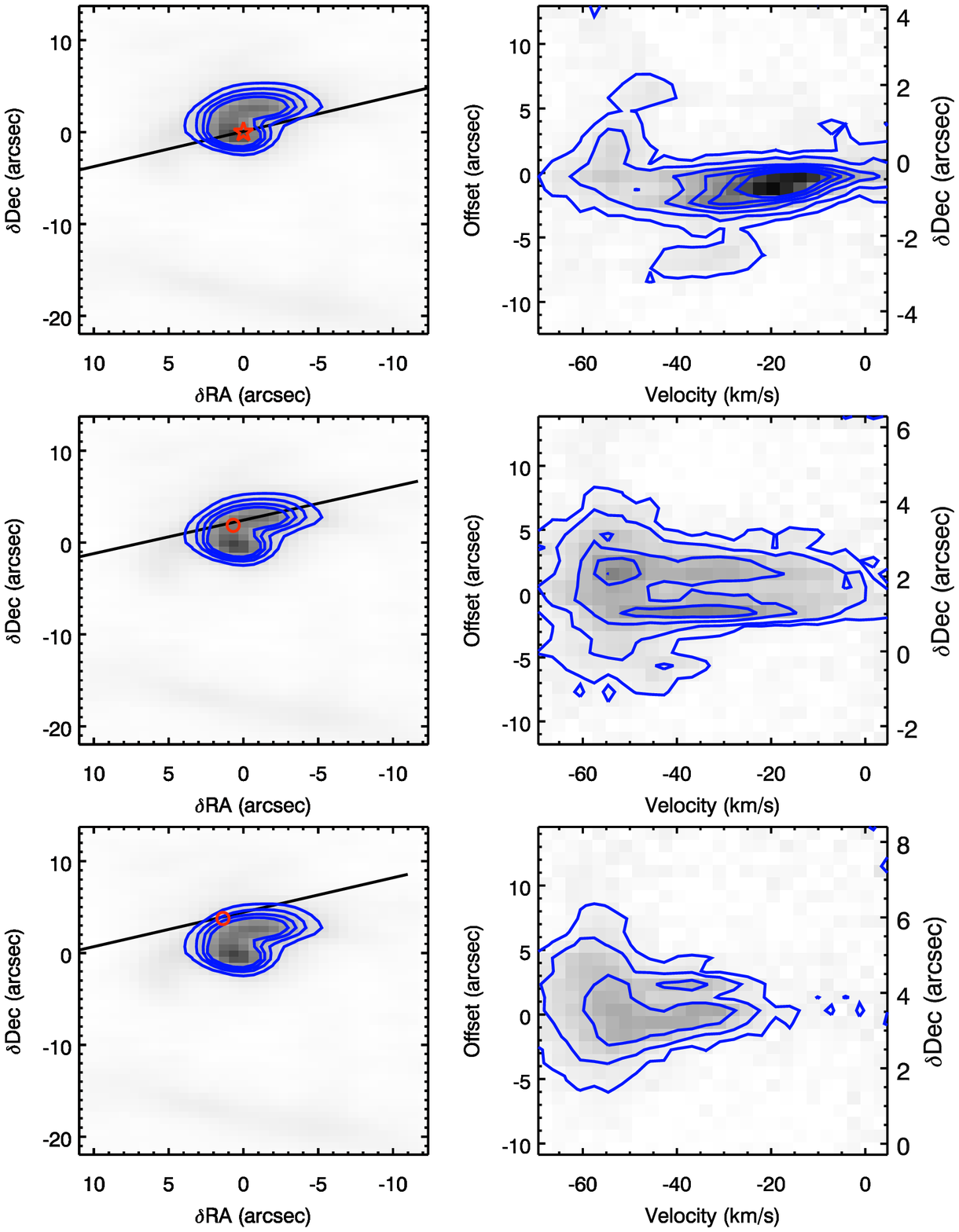,width=0.9\textwidth,angle=0}
       }
 \caption{The same as Figure \ref{f:h2_par} except for the cut 
 perpendicular to the symmetry axis. In the second and third panels, 
 the cuts are in the 2\arcsec\ and 
 4\arcsec\ northeast (top left) away from the cut in the first panel. }
   \label{f:h2_pen}
 \end{figure*}

On the other hand, the PV diagrams of H2 in Figs.~\ref{f:h2_par} and~\ref{f:h2_pen} 
are fully consistent with predictions of the pressure driven model. 
For the cuts parallel to the symmetric axis, near the star, the ionized gas has the lowest velocity 
relative to nearby neutral gas. When moving away from the apex of the paraboloid, the ionized 
gas has been accelerated due to the pressure gradient. 
For the cuts 
perpendicular to the symmetric axes, both sides of P35 collect the emission from 
ionized gas along the front and back sides of the paraboloid and therefore show large velocities 
spanning a large range, most of which are blueshifted relative to the nearby 
neutral gas (Fig.~\ref{f:h2_pen}).

In summary, our \irtf/TEXES data show more detailed velocity structures  
in the two \ion{H}{2} regions, and provide further evidence that 
the scenarios proposed in 
~\citet{don15} are correct.
 
 \subsection{A new star cluster in the central 10\arcmin\ of the GC?}\label{ss:d_c}
If H2 indeed is the natal cloud of P35, one would then 
wonder whether or not P35 has formed alone or, as is more likely, is part of 
a cluster.  We therefore examine in detail 
the existing near-IR photometry data to study the stellar number density distribution 
and the CMD of the stars near P35.

\begin{figure*}[!thb]
  \centerline{    
       \epsfig{figure=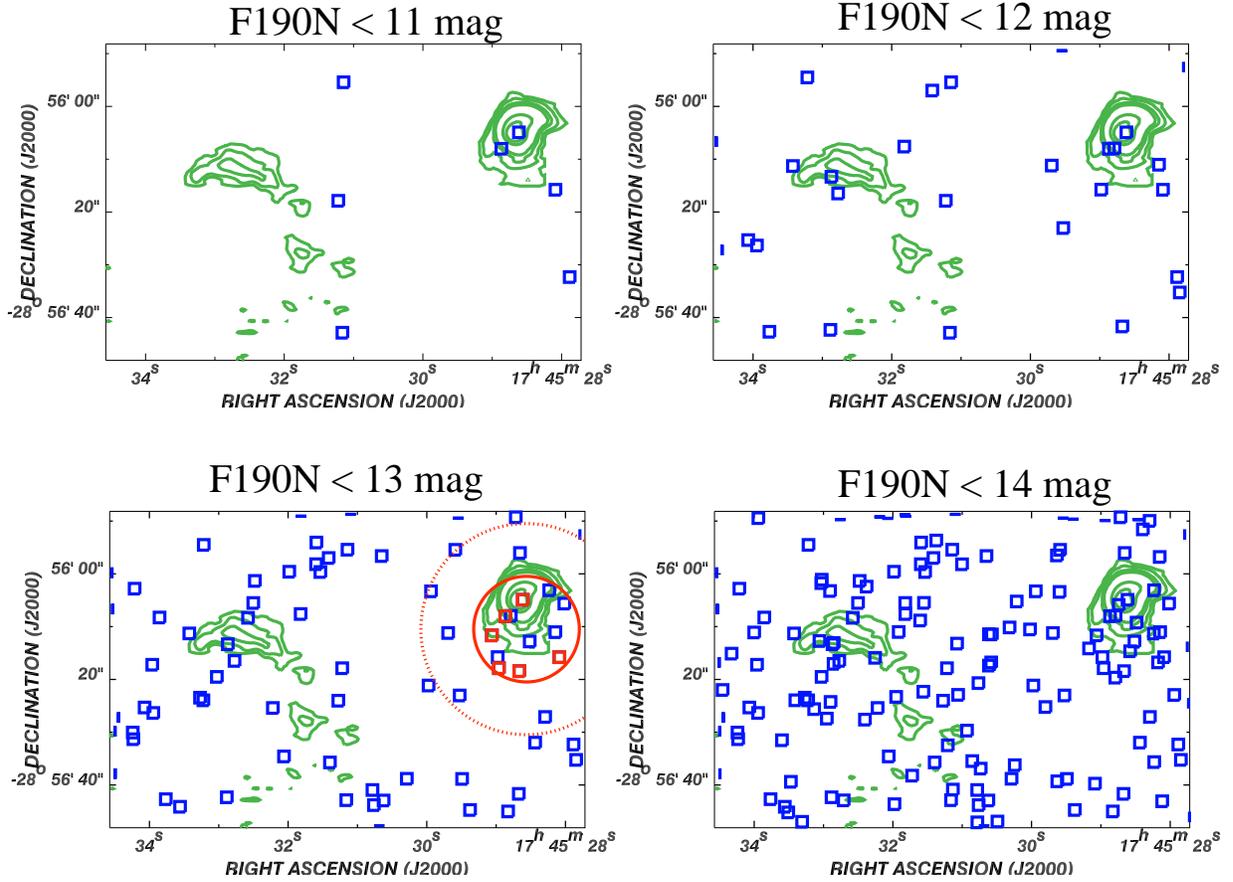,width=1.0\textwidth,angle=0}
       }
 \caption{The spatial distributions of stars (squares) 
 brighter than the magnitude, as labeled at the top of each 
 panel. The field-of-view of 
 these images and the green contours are the same as those 
 of Fig.~\ref{f:2d_image}. The solid and dashed red circles in the lower left panel 
mark the regions of 10\arcsec\ and 20\arcsec\ radii centered on H2. Six 
stars, colored in red within the 10\arcsec\ radius have 
$H$-$K_s$ colors, similar to P35 (see Fig.~\ref{f:cmd}).}
 \label{f:num_dis}
 \end{figure*}

Fig~\ref{f:num_dis} compares the spatial distribution of stars with F190N magnitudes greater than 
11, 12, 13 and 14 from our \hst/NICMOS survey\footnote{We do not use the \vlt/HAWKI observation 
at the $K_s$ band here, because bright stars saturate.}. Those dimmer than 
14 mag should mostly be Red Clump stars 
(Figure 14 in~\citealt{don11}). Although it is difficult to see a clear 
concentration of stars brighter than 12 mag due to the small stellar number density, we can see 
an enhancement 
of stellar number density near the H2 for stars with F190N $<$ 13 mag, but not in H1. 
These stars are 
distributed mostly in the 
direction away from the -30-0 {\rm km s$^{-1}$} molecular 
cloud, which is in the north of 
H1 and H2.  
In order to further quantitatively demonstrate this enhancement, 
we smoothed the stellar number density 
map for stars with F190N $<$ 13 mag by a Gaussian kernel 
with a full width half maxim 
of 20\arcsec . The stellar number densities in H1 and H2 are 0.017 
and 0.022 arcsec$^{-2}$, which are 1.7 and 3 sigma above the mean 
value in the field-of-view of Fig.\ref{f:2d_image}. This confirms that there are 
more stars near H2. We examined the extinction 
map (A$_{F190N}$) given in~\citet{don11} to explore the possibility whether the number 
density fluctuation is introduced by the spatial variation of the extinction. In our 
field-of-view, the median and 68\% of the extinction, A$_{F190N}$, is 3.11 mag 
and 0.27 mag. On the other hand, the A$_{F190N}$ 
within 10\arcsec\ of H2 and P35, A$_{F190N}$ are 
3.07$\pm$0.36 and 3.2 mag (derived from the $J$-$K$ color in~\citealt{don12}), which is not systematically 
lower than the average value of the whole field-of-view. Therefore, we conclude that 
the spatial variation of extinction is not responsible for the higher stellar number density in H2. 

\begin{figure*}[!thb]
  \centerline{    
       \epsfig{figure=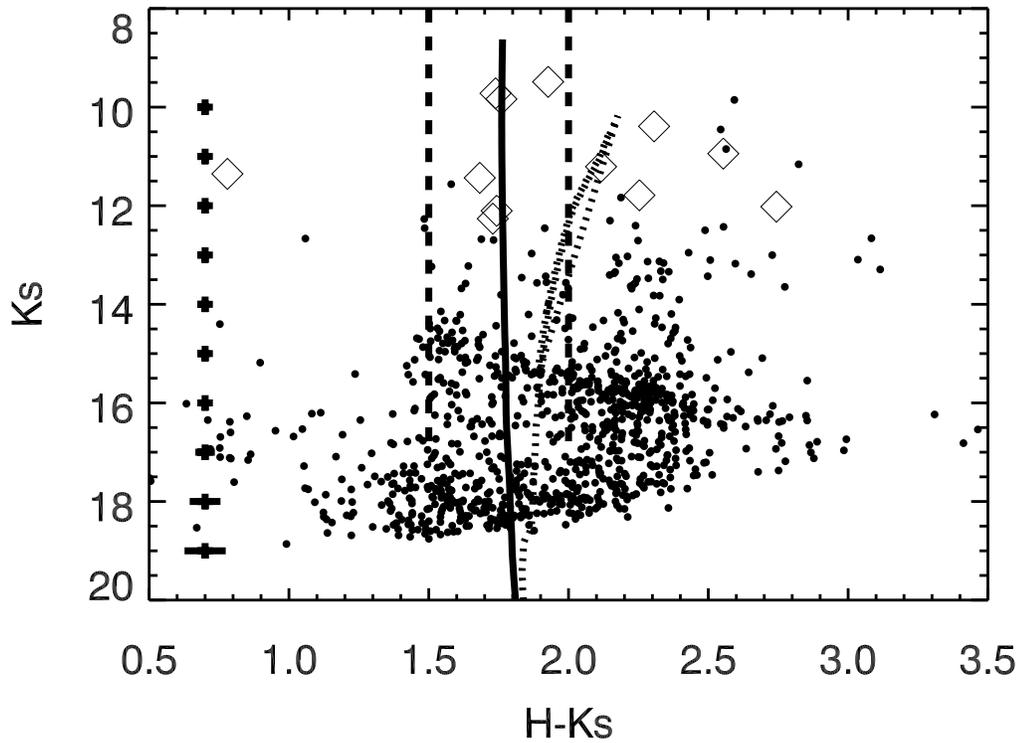,width=1.0\textwidth,angle=0}
       }
 \caption{The CMD of the stars with available 
 \vlt/HAWKI magnitudes in the central 20\arcsec\ of H2. Those  
12 stars with F190N magnitudes 
 brighter 13 within the central 10\arcsec\ are marked as diamonds. 
 The error bars (heavy black crosses on the left side) show 
 typical uncertainties in $K_s$ and $H$-$K_s$. The solid and the dotted lines represent 
 the 2 Myr old Genova and 5 Gyr old Padova isochrones with 
 solar metallicity, respectively. The two vertical dashed lines enclose 
 the color range, which we use to select the six stars (red squares in the lower right 
 panel of Fig.~\ref{f:num_dis}) with $H$-$K_s$ similar to P35. }
 \label{f:cmd}
 \end{figure*}

In Fig.~\ref{f:cmd}, we plot the  $H$$-$$K_s$ vs. $K_s$ CMD of the stars within the central 
20\arcsec\ of H2. 
Unfortunately, most of the stars (76\%) do not have 
available $J$ band magnitudes; $J$$-$$K_s$ would allow for a better stellar population 
analysis. Six of the 12 stars with available F190N magnitude within the 
central 10\arcsec\ of H2 have similar $H$-$K_s$ colors.~\citet{don15} estimate the 
age of P35 to be roughly 2 Myr. Therefore, we overlay a Geneva stellar 
isochrone~\citep{eks12} with solar metallicity and an age of 2 Myr 
on the CMD of Fig.~\ref{f:cmd}. 
The distance modulus (8 kpc) and the foreground extinction derived from the color of P35 
($A_{Ks}$=2.45,~\citealt{don12}) are included in the calculation of 
the isochrone. The six stars are indeed very close to the 2 Myr old isochrone, 
and are certainly bluer than the similar isochrone of an age 5 Gyr. This result 
should be robust because of the small color uncertainty of these bright stars.  If they were evolved low-mass 
stars, their colors would suggest an extinction $A_{Ks}$=0.3 mag less than that derived for P35. 
Such a decreased extinction would not be consistent with the nearly constant ratio of the [\ion{Ne}{2}] and Paschen-$\alpha$ intensities across 
H2. Therefore, it is unlikely that these stars are evolved low-mass 
stars that have similar colors due to lower extinction. 

\subsection{The impact on the star formation in the GC}\label{ss:d_s}
How would our new star cluster fit into the big 
picture of the star formation process in the GC? 

Recently,~\citet{ste16} reports a new cluster in the Sickle Nebula, which  
includes one WN6 star (WR102c), one O7? star 
and three B stars. The WN6 star and the other four early-type stars have similar 
radial velocities, which suggests that they could be bound in a cluster 
of total stellar mass $\sim$ 1000 M$_{\odot}$. Because this cluster is 
close to the Quintuplet cluster ($\sim$ 50\arcsec , 2 pc), they may be 
related to the same starburst activity. 

Similarly, our cluster also has only one evolved 
star, P35, an O If star, which will eventually become a WN star~\citep{cro07}. 
Therefore, it is not surprising that the star would be associated with 
a possible cluster with a similar total stellar mass to the new cluster in the Sickle Nebula. 
However, the H2 cluster is far away from any other known ones, especially the three known
 massive clusters (e.g. P35 is $\sim$7.5\arcmin\ and 5\arcmin , i.e. 18 pc and 12 pc, in projected 
 distances away from the Arches and Central clusters). 
 Therefore, P35  
represents a small cluster with at least one massive star 
formed in relative isolation.

\begin{figure*}[!thb]
  \centerline{    
       \epsfig{figure=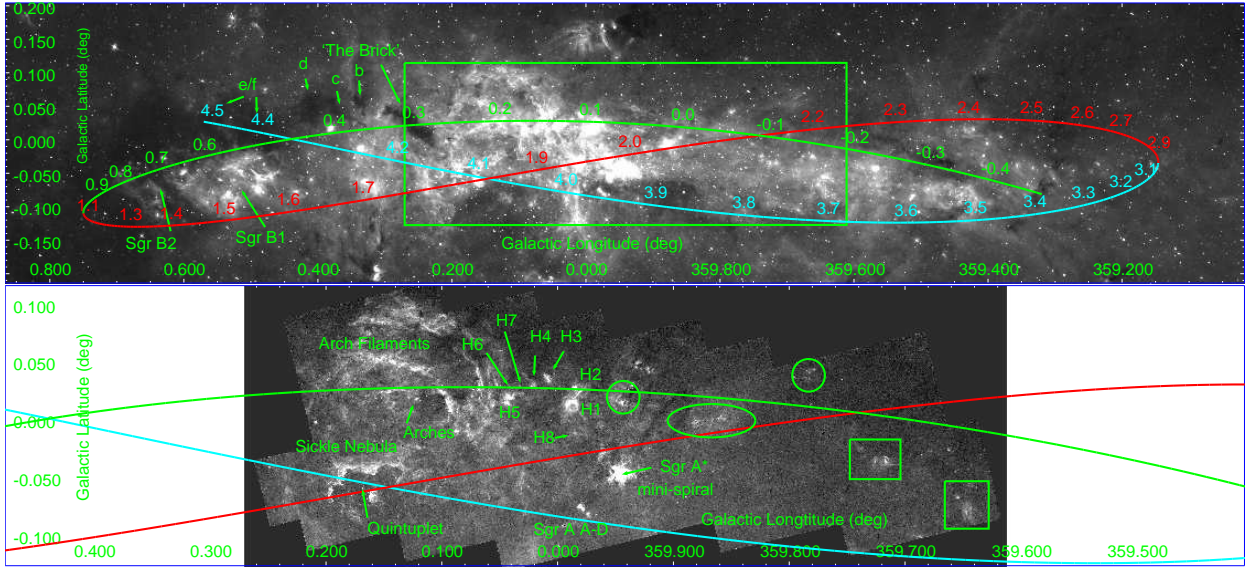,width=1.0\textwidth,angle=0}
       }
 \caption{The hypothetical orbit of various molecular clouds in the CMZ, as proposed by~\citet{kru15}, 
 overlaid on the \spitzer/IRAC 8.0 $\mu$m (top) 
 and \hst/NICMOS Paschen-$\alpha$ images (bottom). The green box in the top panel outlines the field-of-view 
 of the Paschen-$\alpha$ image. The numbers marked above the orbit in the top panel indicates 
 how long individual molecular clouds have passed by Sgr A* in the pericentre (the point closest to Sgr A*) 
 near the `Brick' (There are three pericentre in the orbit of molecular clouds given in~\citet{kru15}); 
 the minus symbol means the 
 time before the passage. Several molecular clouds and two star formation regions, Sgr B1 and Sgr B2, have 
 been marked in the top panel. In the bottom panel, \ion{H}{2} regions 
 in the GC are labeled, such as 
 H1-8~\citep{yus87}. In addition, new \ion{H}{2} regions, identified from our \hst/NICMOS 
 Paschen-$\alpha$ survey are outlined in an ellipse, two circles and boxes.}
 \label{f:orbit}
 \end{figure*}

~\citet{don15} suggest that the star formation processes in H2 could have been triggered 
when the -30-0 {\rm km s$^{-1}$} molecular cloud passed near Sgr A*~\citep{lon13}
 about 2-2.5 Myr ago. According to this scenario, at this point, H2 would be on the back side of the orbit 
proposed by~\citet{mol11}, assuming a period of $\sim$3 Myr. Since~\citet{mol11}, a new orbit for the 
molecular clouds has been proposed by~\citet{kru15}, which is  
overlaid on the \spitzer/IRAC 8.0 $\mu$m image~\citep{sto06} 
and \hst/NICMOS Paschen-$\alpha$ image in Fig.~\ref{f:orbit}. 
The green and cyan parts of the orbit are in the front side, while 
the red portion is the back side. In this model, H2 falls on 
the front orbital side (green line) and is about 0.1 Myr beyond the point of pericentre passage. 
For comparison, the well known `Brick' passed the pericentre point 
$\sim$0.3 Myr ago. This is apparently inconsistent with 
P35 having evolved to an O If star. It thus seems that H2 is in 
conflict with the star formation sequence between the `Brick' and Sgr B2. 
We consider two alternative possibilities: 
1) H2 is not on the front side of the orbits (green line), but the back side (red line). 
The green and red parts of the orbit intersect near H2 in the projection. 
They have similar radial velocities (Figure 4 in~\citealt{kru15}). Therefore, 
it is possible that P35 is associated with the molecular cloud on the back side of the orbit, which 
passed by Sgr A* about 1.6 Myr ago. 2) The formation of the H2 cluster 
was triggered in an even earlier pericentre passage, i.e. 2.1 Myr ago 
(see Figure 6 of~\citealt{kru15}). 

Besides H2, several well-known \ion{H}{2} regions, as well as newly discovered ones~\citep{wan10}, fall 
onto the orbit of molecular clouds and could represent the star formation 
induced by their pericentre passages. First, we consider H3, H4, H6 and H7, 
among the eight \ion{H}{2} regions (H 1-8) identified by 
\citet{yus87} in the region 
between the Arches cluster and Sgr A*.~\citet{don15} show the presence of   
single evolved massive star in each center of H5 and H8. 
They find that the radial velocities of these stars are significantly different from 
those of nearby ionized and neutral gas. This indicates that the stars 
are likely interlopers and did not form {\it{in-situ}}. In the other four 
\ion{H}{2} regions, H3, H4, H6 and H7, no evolved massive stars have been found, 
although the possibility exists that there are still embedded  
massive main-sequence stars in these \ion{H}{2} regions. 
The massive main-sequence stars normally have hydrogen or helium absorption 
lines. Therefore, they cannot be identified by their Paschen-$\alpha$ emission 
lines in the work by~\citet{don12}. Another possibility is that the hydrogen emission 
lines from the central evolved massive stars are diluted by the emission 
from the surrounding dust, like the dusty WC stars~\citep{don12}. 
Furthermore, H6 and H7 with size of $\sim$0.1 pc could be ultracompact \ion{H}{2} regions. 
Compared to H2, the embedded stars could be less massive and therefore 
do not have enough ionization luminosities nor have they had 
time to disperse the surrounding dense molecular gas. 
Second, let us consider several new \ion{H}{2} regions identified by 
our \hst/NICMOS Paschen-$\alpha$ survey in the region Galactic West to Sgr A* (see the  
circles, ellipse and boxes in the bottom panel of Fig.~\ref{f:orbit}). The 
detailed Paschen-$\alpha$ image of the \ion{H}{2} region enclosed by the   
ellipse can be found in Figure 3 of~\citet{wan10}. All of these regions contain 
evolved massive stars, such as WN, Ofpe/WN9, WC9d, B supergiant, 
and their candidates~\citep{mun06,cot99,mau10a,mau10b,don12}. Some of them are close to 
the molecular cloud orbit, which include those marked by the ellipse and the two circles. 
Therefore, they may be molecular clouds that have experienced 
star formation induced by the pericentre passage around  Sgr A*. 
The two regions marked by the boxes are somewhat off the orbit and may have different origins. 
In conclusion, we have demonstrated that star clusters 
associated with the \ion{H}{2} regions, aided by the CMD analysis could help us 
constrain their ages and the star formation history in the GC.  

\section{Summary}\label{s:summary}
In this work, we have mapped the radial velocity distribution of 
of ionized gas in two \ion{H}{2} regions 
in the GC, H1 and H2, using our new \irtf/TEXES observations 
of the [\ion{Ne}{2}] 12.8 $\mu$m line. We have also examined the spatial 
distribution and color magnitude diagram of an apparent stellar concentration in and around H2. 
We summarize our results below:

\begin{itemize}
\item The [\ion{Ne}{2}] emission morphologically follows the Paschen-$\alpha$ emission 
well, indicating both are from the ionized gas produced by massive stars at the surface of adjacent 
molecular clouds. 

\item The intensity ratio between Paschen-$\alpha$ and [\ion{Ne}{2}] line near the bright 
Paschen-$\alpha$ region in H1 is larger than that in the surrounding region, indicating 
a local extinction drop of $\sim$0.56 mag in the K$_s$ band.

\item The velocity structures of H1 and H2 can be well explained by the bow-shock 
and pressure driven models, respectively. 

 \item The stellar concentration around H2 shows a
 color magnitude diagram that is consistent with the presence of a 2 Myr old stellar 
 cluster, as hypothesized previously~\citep{don15}. 
 
 \item The location of H2 on the orbit of a molecular cloud chain, as proposed by~\citet{kru15}, 
 which indicates that the formation of this cluster may be triggered by a close passage of 
 its parent cloud around Sgr A* $\sim$ 2 Myr ago. 
  
\end{itemize}

\section*{Acknowledgments}
We thank the anonymous referee for a thorough, detailed, and
constructive commentary on our manuscript.
The research leading to these results has received funding from the European Research Council under the European Union's Seventh Framework Programme (FP7/2007-2013) / ERC grant agreement n° [614922]. 
F N-L acknowledges financial support from a MECD predoctoral contract, code FPU14/01700. 
This work uses observations made 
with the NASA/ESA Hubble Space Telescope and 
the data archive at the Space Telescope Science Institute, which is 
operated by the Association of Universities for Research in Astronomy, 
Inc. under NASA contract NAS 5-26555. 
This work is also based on observations collected at the European Organisation 
for Astronomical Research in the Southern 
Hemisphere under ESO programme(s) 195.B-0283(A).

\end{document}